\documentclass{webofc}
\usepackage[varg]{txfonts}   
\usepackage{booktabs}

\newcommand{\be}{\begin{equation}}
\newcommand{\ee}{\end{equation}}

\begin{document}
\title{High-Energy Cosmic Rays from Radio Galaxies}

\author{\firstname{Bj\"{o}rn} \lastname{Eichmann}\inst{1}\fnsep\thanks{\email{eiche@tp4.rub.de}} 
}

\institute{Ruhr Astroparticle and Plasma Physics Center (RAPP Center), Ruhr-Universit\"at Bochum, Institut f\"ur 
Theoretische Physik IV/ Plasma-Astroteilchenphysik, 44780 Bochum, Germany}

\abstract{%
A detailed investigation of radio galaxies has recently stressed these sources as the possible origin of the cosmic 
rays observed above $3\,\text{EeV}$.
Here, the relevance of this model at energies below $3\,\text{EeV}$ is investigated. 
So, it is shown that the average contribution of radio galaxies can accurately explain 
the observed CR flux between the second knee and the ankle in the case of a strong source evolution. However, the model 
cannot provide the increasing heaviness and variance at energies $\lesssim 
1\,\text{EeV}$ of the observed chemical composition.
In addition, it is exposed that the resulting variance of the chemical composition at Earth shows also at higher 
energies a clear disagreement with the observations, indicating that the compositional contributions by Centaurus A and 
Cygnus A need to be less different.
}
\maketitle
\section{Introduction}
\label{intro}
The origin of the High-Energy Cosmic Rays (HECRs) is still one of the great enigmas of modern astrophysics.
From observatories like the Pierre Auger Observatory (Auger) and the Telescope Array (TA) experiment at the 
highest energies as well as KASCADE, KASCADE-Grande and a few other detectors at lower energies, there are basically 
three main observational characteristics, that describe our current knowledge of the HECRs:
\begin{enumerate}
 \item[(1.)] The energy spectrum, which changes at about $0.4\,\text{EeV}$ --- the so-called second knee --- to a 
stepper power-law distribution with a spectral index of about 3.3 and flattens above about $3\,\text{EeV}$ 
--- the so-called ankle --- to a spectral index of 2.6 and a sharp flux suppression above about 
$10^{19.5}\,\text{eV}$ \cite{0954-3899-34-10-R01, Abbasi:2007sv, 2010PhLB..685..239A}. 
 \item[(2.)] The chemical composition, that shows a decrease of the fraction of heavier elements between about
$10^{17}\,\text{eV}$ and $10^{18.3}\,\text{eV}$ as well as an increase at energies $>10^{18.3}\,\text{eV}$ 
\cite{Abraham:2010yv, KAMPERT2012660, Aab:2014aea, ObservatoryMichaelUngerforthePierreAuger:2017fhr}. 
 \item[(3.)] The arrival directions, that are usually expressed in terms of the multipoles of their spherical 
harmonics. Here, Auger recently reported a $5\sigma$ detection of a dipole with an amplitude 
of $\approx 6.5\%$ at energies $>8\,\text{EeV}$, while higher-order multipoles as well as the multipoles at lower 
energies are still consistent with isotropy \cite{Aab:2017tyv}.
\end{enumerate}
A likely source candidate of those extremely energetic particles are radio galaxies due to their powerful acceleration 
sights within the jets, as already noted by Hillas in 1984 \cite{Hillas:1985is}.
A recent study by Eichmann et al. \cite{1475-7516-2018-02-036} --- hereafter referred to as E+18 --- investigated in 
great detail the contribution of CRs above $3\,\text{EeV}$, called ultra-high-energy cosmic rays (UHECRs), by radio 
galaxies. This model comprises three components: 
(a) a 3D structure of radio galaxies and EGMF within a radius of 120 Mpc; (b) a continuous source function (CSF) 
derived from a luminosity function of radio galaxies for the contributions from beyond 120 Mpc; (c) a contribution of 
the powerful radio galaxy Cygnus A, which is near the magnetic horizon and is expected to deliver the dominant 
contribution to UHECRs due to its extreme radio power and brightness. 
All simulations have been carried out with CRPropa3 \cite{1475-7516-2016-05-038}, including deflections inside the 
Galaxy according to the magnetic field model of Jansson \& Farrar \cite{2012ApJ...757...14J}. 
So, E+18 draw some solid conclusions on the UHECR contribution by radio galaxies:
\begin{enumerate}
 \item[(i)] The average contribution of all radio galaxies cannot explain both spectrum and composition of UHECR. 
 \item[(ii)] The spectrum and composition of UHECR can be explained by the two brightest radio galaxies in the sky: 
Cygnus A and Centaurus A. Here, Cygnus A needs to provide a solar-like composition, while Centaurus A needs to have a 
heavy composition, with an iron fraction comparable to protons at a given cosmic-ray energy. Further, both sources need 
a cosmic ray load significantly above the average of the bulk of radio galaxies. 
 \item[(iii)] Also the anisotropy constraints at energies ${>}\,8\,$EeV are satisfied, if the UHECRs from Cygnus A are 
significantly deflected by the EGMF.
\end{enumerate}

In this paper, the model of E+18 is used to check additional aspects of the observational data that have not been taken 
into account so far. First, the Sect.~\ref{sec-1} summarizes the physics of the E+18 model. Subsequently, in 
Sect.~\ref{sec-2} the best-fit models of E+18 is applied in order to discuss the CR contribution by the CSF below the 
ankle as well as the variance of the mean logarithm of the mass number.

\section{The E+18 model}
\label{sec-1}
As shown in great detail in E+18, the established correlation between the jet power $Q_{\rm jet}$ and the extended radio 
luminosity $L_{1.1}$ from Willott et al.\ \cite{1999MNRAS.309.1017W} gives for the minimal energy condition 
\cite{1970ranp.book.....P} a good estimate of the cosmic ray luminosity 
\be
Q_{\rm cr}\simeq \frac47 Q_{\rm jet} = 1.3\times 10^{42}\,g_{\rm cr}\, \left({L_{1.1} \over L_*} 
\right)^{6/7}\,\frac{\text{erg}}{\text{s}}\,,
\label{Qcr}
\ee
and the maximal rigidity
\be
\hat{R}(Q_{\rm cr}) = g_{\rm acc} \sqrt{Q_{\rm cr}/c}\,,
\label{Rmax}
\ee
where $L_* \approx 4.9\times 10^{40}\,\text{erg/s}$ denotes a characteristic luminosity according to Mauch and 
Sadler \cite{2007MNRAS.375..931M}.
In doing so, we suppose that the total power in CRs is significantly higher than in relativistic electrons. 
From the uncertain efficiency of converting jet internal energy into observable radio luminosity, as well as the 
uncertain details of the acceleration process one obtains the dimensionless coefficients 
\be
1 \lesssim g_{\rm cr} \lesssim 50\quad{\rm and}\quad 
0.01 \lesssim g_{\rm acc} \lesssim 0.5\;.
\ee

Introducing different nuclei species $i$ with charge number $Z_i$ and an abundance $f_i$, the total cosmic ray power per 
charge number yields
\be
Q_{{\rm cr},i}\equiv Q_{\rm cr}(Z_i)=f_i\,Z_i\,Q_{\rm cr}/\bar Z\,.
\label{eq:totCRpowerPerZi}
\ee
Here, a simple abundance relation $f_i = f_{\odot} Z_i^q$ has been suggested, where $f_{\odot}$ denotes solar 
abundances, so that the heaviness of the initial composition can be increased by a single parameter $q$. 

Apart from the individual treatment of the sources from the catalog of van Velzen et al.\ 
\cite{2012A&A...544A..18V}, the sources beyond a distance of 120 Mpc from Earth, except for Cygnus A, are 
treated by a CSF in a 1D simulation. Here, only the impact of cosmic photon targets is 
taken into account due to the lack of a known EGMF. So, the local CSF, $\Psi_{i,0}(R)$, is derived as
\begin{equation}
\Psi_{i,0}(R) \equiv {\mathrm{d}N_{\rm cr}(Z_i) \over \mathrm{d}V \mathrm{d}R\,\mathrm{d}t} = \int_{\check Q_{\rm 
cr}}^{\hat Q_{\rm cr}} S_i\big(R,\hat R(Q_{\rm cr})\big)\,{\mathrm{d}N_{\rm RG} \over \mathrm{d}V\,\mathrm{d}Q_{\rm 
cr}}\,\mathrm{d}Q_{\rm cr}\,,
\label{CRsourceRateDensity}
\end{equation}
using the local radio luminosity function, $\mathrm{d}N_{\rm RG}/(\mathrm{d}V\,\mathrm{d}Q_{\rm cr})$, by Mauch 
and Sadler \cite{2007MNRAS.375..931M} and 
the CR spectrum, $S_i\big(R,\hat 
R(Q_{\rm cr})\big)$, of element species $i$ with charge number $Z_i$, emitted by a source with total cosmic ray power 
per charge number, $Q_{{\rm cr},i}$, up to a maximal rigidity $\hat R(Q_{\rm cr})$.
The limits of integration are the smallest, $\check Q_{\rm cr}$, respectively largest, $\hat Q_{\rm cr}$, CR powers that
need to be considered. 
To solve the integral analytically, as shown in E+18, a sharp cutoff of the individual source spectra at $\hat R(Q_{\rm 
cr})$ is supposed. 
The function $\Psi_{i,0}(R)$ is the local continuous source function as it is derived from a radio luminosity function 
determined in the local universe ($z<0.3$). To extend it to larger redshifts, the approximation
\be
\Psi_i(R,z) \approx \Psi_{i,0}(R)\,(1+z)^{m-1}\,,
\label{eq:SourceFunctionEvolution}
\ee
is used with a source evolution index $m\in[3,5]$.
The analytical solution of the CSF (\ref{eq:SourceFunctionEvolution}) shows a spectral break at a critical rigidity $R_* 
= g_{\rm acc} \sqrt{g_{\rm cr}\,Q_* / c} \approx 2\times 10^{18}\,g_{\rm acc}\,\sqrt{g_{\rm cr}}\,\text{V}$. Thus, the 
spectral behavior of the CSF yields $\Psi_i(R<R_*,z)\propto R^{-a}$, where $a$ denotes the spectral index of the 
individual sources, and at high rigidities 
\be
\Psi_i(R>R_*,z)\propto \begin{cases}
                        R^{-3}\,,\quad &\text{for }\, a\leq 2,\\
                        R^{-1-a}\,,\quad &\text{for }\, a> 2.
                       \end{cases}
\ee
As already stressed in E+18, this spectral behavior impedes an explanation of the CR data above the ankle by the CSF.
\section{New Prospects}
\label{sec-2}
In order to investigate the observational data below the ankle, the simulations of E+18 are performed again including 
energies down to $0.1\,\text{EeV}$. Further, the large compositional gap between oxygen and iron is filled by adding 
silicon to the initial composition of the sources. However, there is no significant impact of silicon on the previously 
obtained results. 
\subsection{CR Flux above the second knee}
\begin{figure}[tbh]
  \centering
    \includegraphics[width=0.49\textwidth]{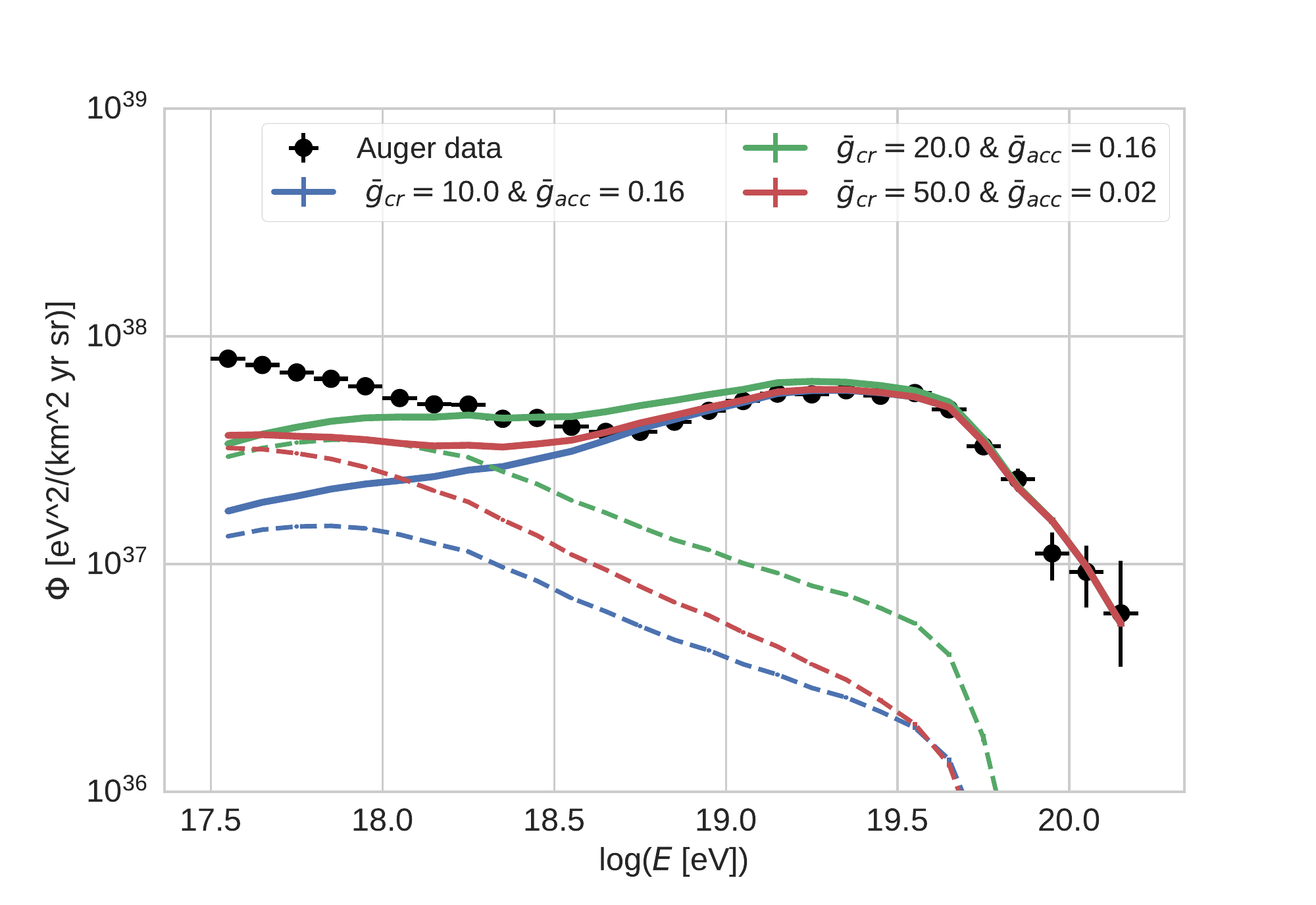}
    \includegraphics[width=0.49\textwidth]{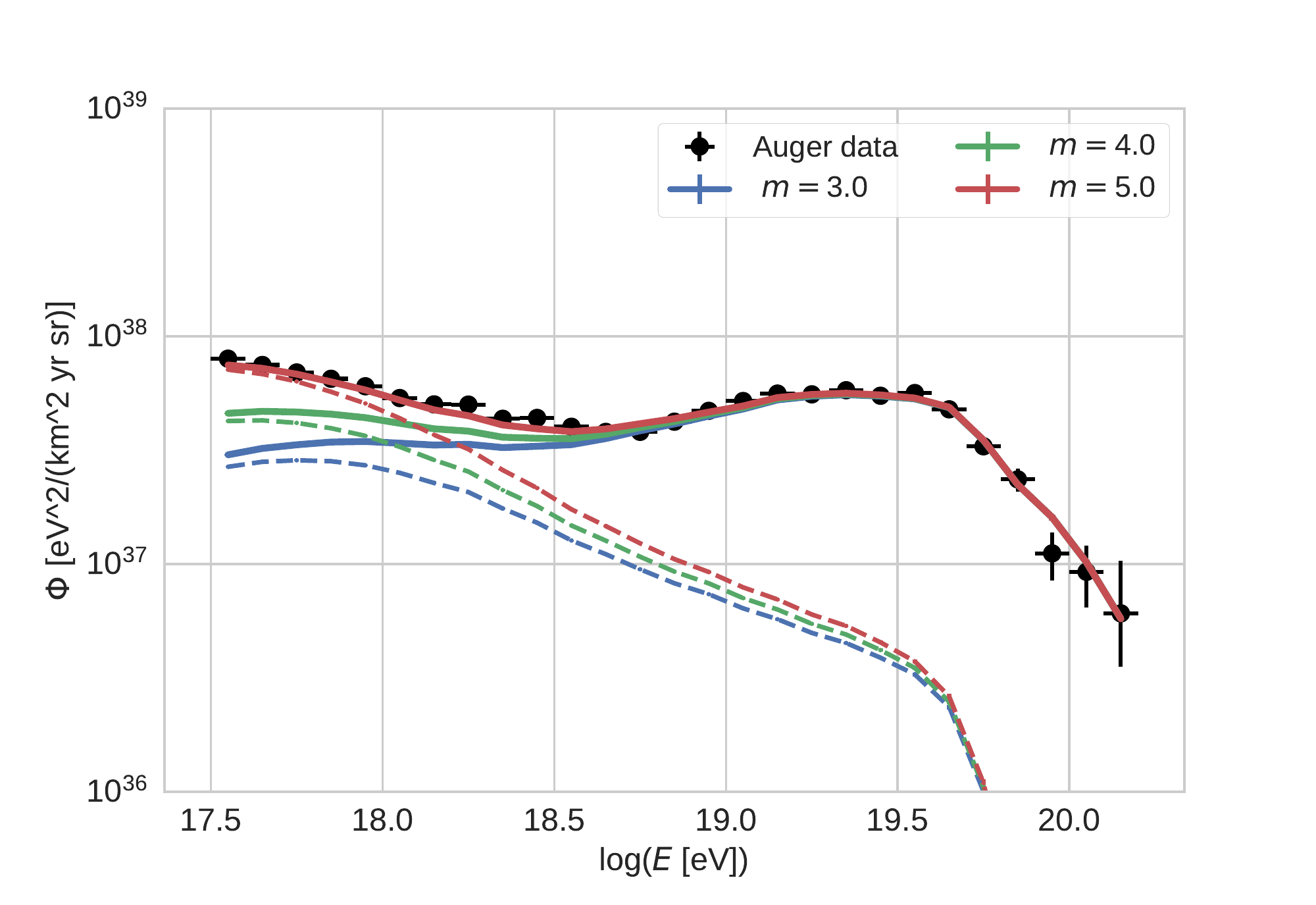}
\caption{Contribution of the CSF (dashed line) on the total flux (solid line). The left figure shows three
different parameter configurations in the case of a source evolution index $m=3$. The right figure displays three 
different source evolution indexes, while the other parameters are chosen according to table \ref{FitPar1}.}
\label{CSF_cont}
\end{figure}

Based on the best-fit scenario from E+18 the impact of the CSF below the ankle is investigated using individual 
parameter values for Centaurus A and Cygnus A and considering the hadronic interaction model of EPOS-LHC.
First the influence of the source evolution parameter is analyzed, yielding an accurate explanation of 
the CR flux above the second knee by the E+18 model in the case of a source evolution index $m\sim5$. The 
Fig.~\ref{CSF_cont} shows that only a large 
value of $m$ provides a spectral contribution by the CSF that is steep enough to describe the observed 
dip of the flux at the ankle. Otherwise, Centaurus A and Cygnus A need an initial spectral index $a\ll 1.8$ 
contravening the first-order Fermi acceleration theory, and as a consequence their baryonic 
load needs to become significantly smaller. 

The parameter values from a simple trial and error fitting to the data are given in table \ref{FitPar1} and the 
corresponding CR fluxes are displayed by the red lines in the right Fig.~\ref{CSF_cont}. 
\begin{table}[h!]
\centering
\caption{Best-fit parameters using solar abundances for all sources except for Centaurus A.}
  \begin{tabular}{ c c c c c c c c c}
  \toprule
             $m\quad$ & $a\quad$ & $\bar{g}_{\rm cr}$ & $g_{\rm cr}^{\rm CenA}$ & $g_{\rm cr}^{\rm CygA}\quad$ & 
$\bar{g}_{\rm acc}$  & $g^{\rm CenA}_{\rm acc}$ & $g_{\rm acc}^{\rm CygA}$ & $q^{\rm CenA}$  \\ 
  \midrule
    $5\quad$ & $1.8\quad$ & $18$ & $25$ & $27\quad$ & $0.09$ & $0.18$ & $0.08$ & $2$  \\ 
   \bottomrule
   \multicolumn{9}{l}{\footnotesize Upper index of the parameter indicates the corresponding source (Centaurus A or 
Cygnus A),}\\
   \multicolumn{9}{l}{\footnotesize bar on top of the parameter corresponds to all other sources}\\
\end{tabular}
  \label{FitPar1}
\end{table}
Note, that this result gives rather a proof of principle than the most likely parameter configuration. In addition, it 
is shown that the average distribution of radio galaxies can provide the same --- or even a higher --- 
baryonic load as Centaurus A or Cygnus A in the case of a small acceleration efficiency compared to Centaurus A. 
Further, the critical rigidity $R_*$ significantly decrease with decreasing $\bar{g}_{\rm acc}$ leading to a steeper 
spectral behavior around the ankle as seen in the left Fig.~\ref{CSF_cont}. Thus, more detailed parameter studies are 
needed to give accurate constraints on the CR physics of the CSF.
\subsection{Chemical Composition}
\begin{figure}[tbh]
  \centering
    \includegraphics[width=0.49\textwidth]{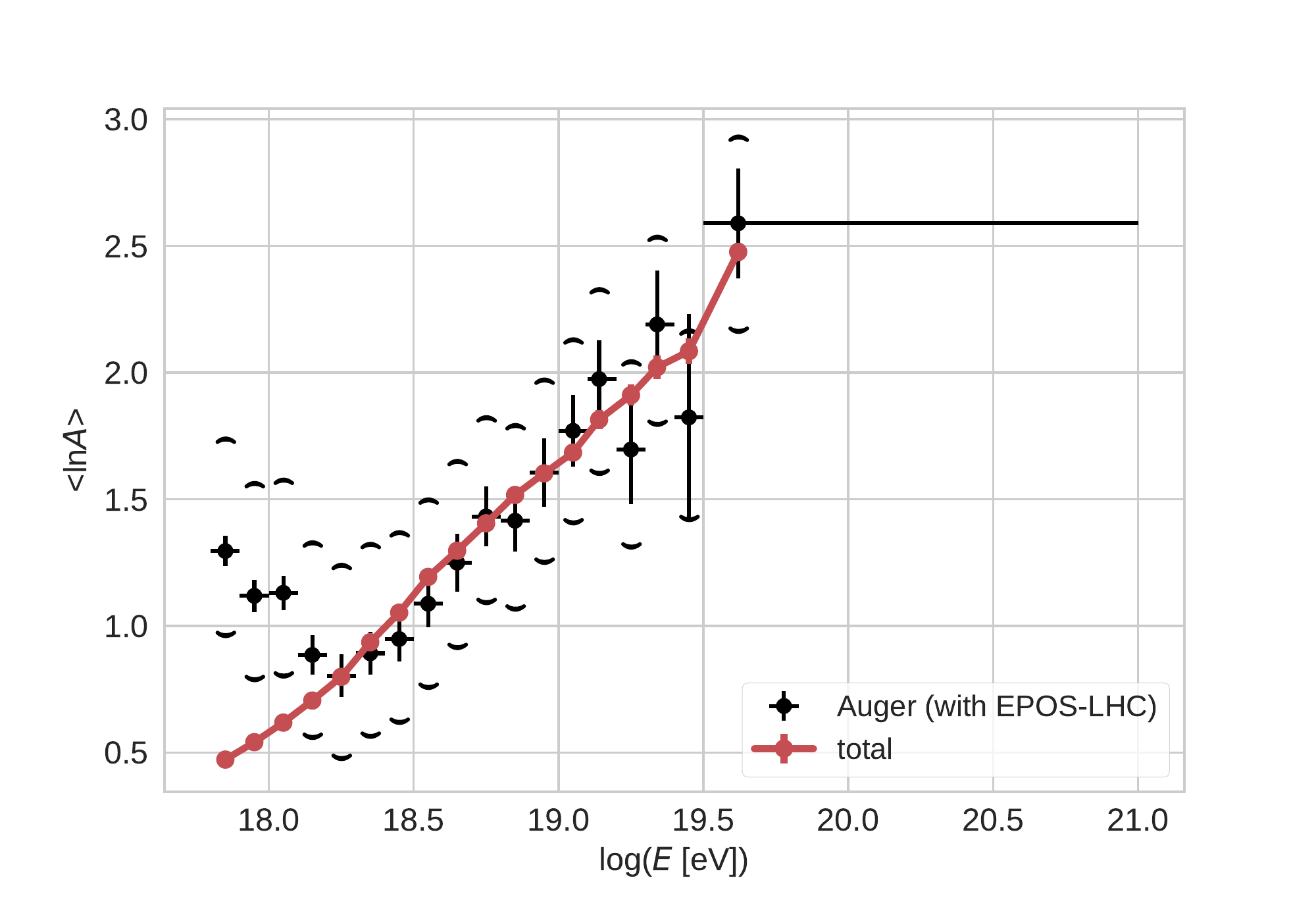}
    \includegraphics[width=0.49\textwidth]{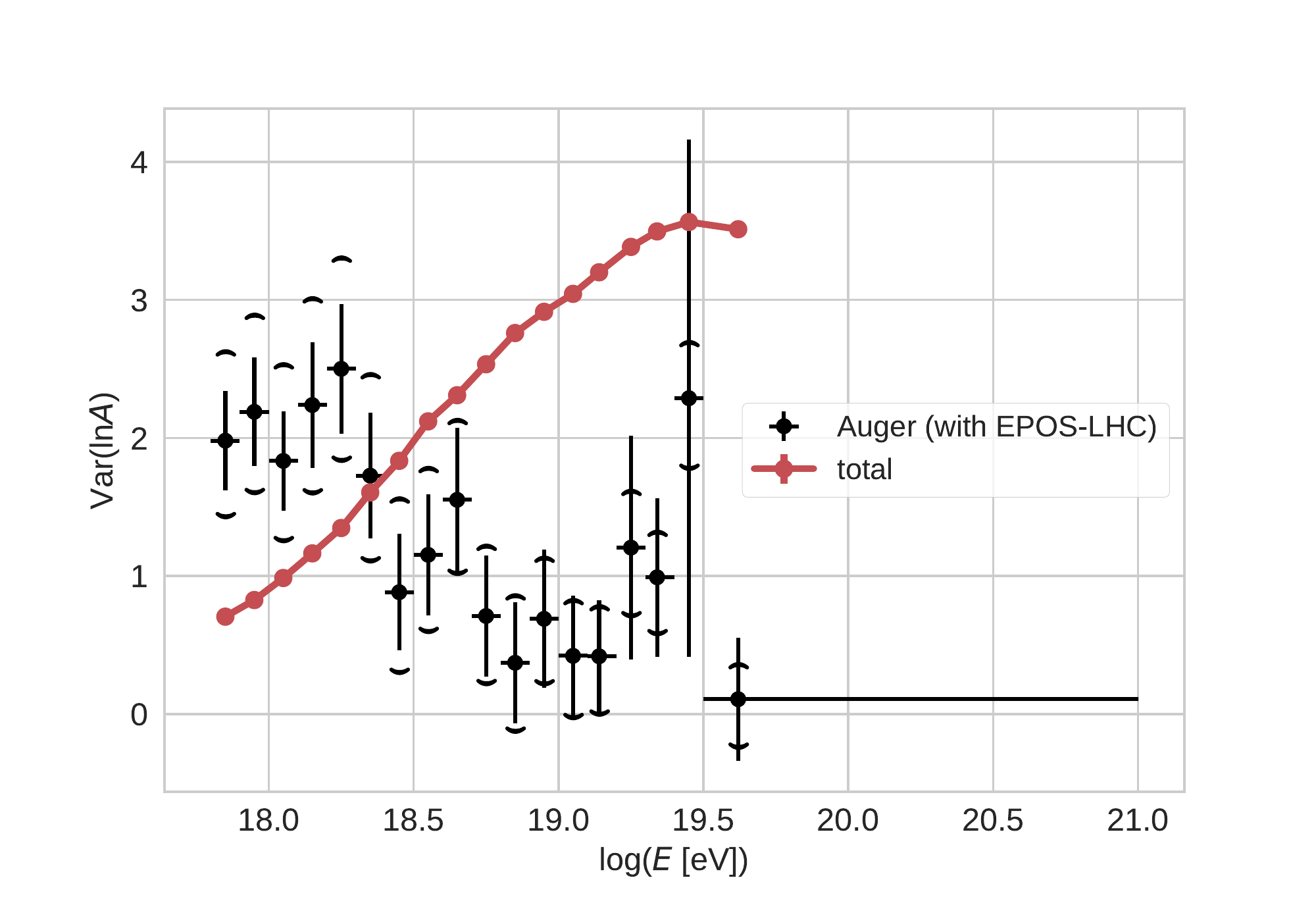}
\caption{$\langle \ln A \rangle$ (left figure) and $\text{Var}(\ln A)$ (right figure) based on the fit parameters 
of table \ref{FitPar1}.}
\label{composition}
\end{figure}

The previously derived fit parameters yield also an excellent 
agreement with the $\langle \ln A \rangle$ data at energies $>10^{18.2}\,\text{eV}$, as shown in the left 
Fig.~\ref{composition}. However, the CSF cannot account for the increase of $\langle \ln A \rangle$ at 
energies $<10^{18.2}\,\text{eV}$. Here, another source contribution with a rather heavy 
composition is needed, as also indicated by the variance of $\ln A$ that is discussed in the following.

So far, the E+18 model has not been tested with respect to the resulting variance of $\ln A$. There are good 
reasons, as the observationally derived $\text{Var}(\ln A)$ values need to be treated with some caution, 
since there are multiple presumptions needed to obtain this observable. Especially an appropriate parametrization of the 
$X_{\rm max}$ distribution is needed, which is usually done by generalized 
Gumbel functions \cite{1475-7516-2013-07-050}. And note that the hadronic interaction model QGSJetII-04 yields negative 
$\text{Var}(\ln A)$ values \cite{Aab:2014aea}, hence, either the QGSJetII-04 model, or the resulting variance analysis 
needs to be defective. 

Keeping this in mind, however, it is clearly shown in the right Fig.~\ref{composition}, that the model cannot explain 
the observed trend of the $\text{Var}(\ln A)$ values. Basically, the mixing of the 
strongly differing compositions of Centaurus A and Cygnus A causes too much variance of the chemical composition at 
Earth. Further, an additional source contribution with a rather heavy initial abundance is needed in order to explain 
the increase in the variance as well as in the mean of the logarithm of the mass number at energies 
$<10^{18.2}\,\text{eV}$. Obviously, a solar-like CSF can not do this job. In total, the compositional data by Auger 
suggest a heavy CR contribution that cuts-off around $1\,\text{EeV}$ and a rather pure chemical composition 
around $10\,\text{EeV}$. Both is not provided by the E+18 model so far.
\section{Conclusion}
The inclusion of observational data and simulations below the ankle demonstrates that the CSF for a 
source evolution index $m\sim 5$ provides an accurate description of the CR flux data between about the second knee 
and the ankle, whereas Cygnus A and Centaurus A appropriately take over at higher energies. Also the $\langle \ln A 
\rangle$ data at energies $>10^{18.2}\,\text{eV}$ is matched perfectly by the model.
However, the comparison of the resulting variance of $\ln A$ with observational data shows a clear disagreement and 
exposes the need for less variance between the initial abundances of Centaurus A and Cygnus A. Up to a certain level, 
this huge difference is a consequence of the simplified abundance relation, which predominantly yields either a proton 
or an iron dominated outflow, and can hardly account for the dominance of the CNO group. 
In total, the E+18 model also provides a promising CR contribution below the ankle, but the chemical composition 
clearly indicates the need for an additional heavy CR source at energies $\lesssim 1\,\text{EeV}$ --- most likely of 
galactic origin. 

\bibliography{references}

\end{document}